\documentclass[doublecol]{epl2} 

\usepackage{amsmath}
\usepackage{amssymb}

\renewcommand{\vec}[1]{{\bf#1}}

\title{Bloch's theory in periodic structures with Rashba's spin-orbit\\interaction}

\shorttitle{Bloch's theory in periodic structures with Rashba's spin-orbit interaction}

\author{S. Smirnov\inst{1} \and D. Bercioux\inst{1,2} \and M. Grifoni\inst{1}}
\shortauthor{S. Smirnov \etal}

\institute{                    
  \inst{1} Institut f\"ur Theoretische Physik, Universit\"at Regensburg, D-93040 Regensburg, Germany\\
  \inst{2} Physikalische Institut, Albert-Ludwigs-Universit\"at, D-79104 Freiburg, Germany
}

\abstract
{
We consider a two-dimensional electron gas with Rashba's spin-orbit interaction and two in-plane potentials superimposed
along directions perpendicular to each other. The first of these potentials is assumed to be a general periodic potential
while the second one is totally arbitrary. A general form for Bloch's amplitude is found and an eigen-value problem for
the band structure of the system is derived. We apply the general result to the two particular cases in which either the
second potential represents a harmonic in-plane confinement or it is zero. We find that for a harmonic confinement regions
of the Brillouin zone with high polarizations are associated with the ones of large group velocity.
}

\begin{document}

\maketitle

\section{Introduction}
It is well known that in a two-dimensional electron gas (2DEG) formed in a semiconductor by an asymmetric confining
potential Rashba's spin-orbit interaction (RSOI) \cite{Rashba} plays an important role. It is very attractive for
applications in electronic devices because the spin-orbit coupling strength can be controlled by an external gate voltage
\cite{Nitta}. Other spin-orbit mechanisms such as Dresselhaus' spin-orbit interaction \cite{Dresselhaus} can be relevant.
In this work, for simplicity, we focus on the effects of RSOI since Dresselhaus' term can be treated in full analogy.
Additionally, when externally enhanced, RSOI may become stronger than other spin-orbit interactions. In this case the
Hamiltonian has the form:
\begin{equation}
\hat{H}^{\text{2D}}_{0}=\frac{\hbar^2\hat{\vec{k}}^2}{2m}-\frac{\hbar^2k_{\text{so}}}{m}\bigl(\hat{\sigma}_{x}\hat{k}_z-\hat{\sigma}_z\hat{k}_x\bigl),
\label{hamiltonian_2D}
\end{equation}
where $k_{\text{so}}$ is the spin-orbit coupling strength and the operator $\hat{\vec{k}}$ is related to the momentum operator
$\hat{\vec{p}}$ as $\hat{\vec{p}}=\hbar\hat{\vec{k}}$. The eigen-states of (\ref{hamiltonian_2D}) have a
two-dimensional spinorial part $\phi^{\text{2D}}_\lambda(\sigma)$ where $\lambda=1,2$ is the eigen-state index called chirality and $\sigma=\pm1$ is the
spin index. The eigen-energies split into two branches $\varepsilon^{\text{2D}}_\lambda(k_x,k_z)$. Systems with such energy spectrum can be
exploited to study spin-dependent transport in different semiconductors, especially in III-V compounds because of the
large values of the spin-orbit coupling strength. As mentioned above they are also used to build both two-dimensional (2D)
and essentially one-dimensional (1D) electronic devices. One such device, called spin transistor, was proposed in Ref.
\cite{Datta} for the case of a quasi-1D system with RSOI. It is obtained from the 2DEG described by
(\ref{hamiltonian_2D}), where by further confinement along, {\it e.g.} the $z$-direction, a quasi-1D wire is formed. These
quasi-1D systems were investigated for the case of a harmonic $z$-confinement \cite{Moroz,Governale} and for an infinite
square well $z$-confinement \cite{Perroni}. In general one can conceive a situation where an arbitrary potential $V(z)$
along the $z$-direction is present. We would like to emphasize that $V(z)$ must not necessarily be a confinement. In this
case the Hamiltonian is written as:
\begin{equation}
\hat{H}^{\text{2D}}_z=\frac{\hbar^2\hat{\vec{k}}^2}{2m}+V(\hat{z})-
\frac{\hbar^2k_{\text{so}}}{m}\bigl(\hat{\sigma}_{x}\hat{k}_z-\hat{\sigma}_z\hat{k}_x\bigl).
\label{hamiltonian_z}
\end{equation}
The Hamiltonian $\hat{H}^{\text{2D}}_z$ assumes that the spin-orbit interaction caused by $V(z)$ is much weaker than RSOI induced
by an asymmetric confinement forming the 2DEG. For a given system this means that the out-of-plane electric field should
be much stronger than the in-plane one.

In systems described by (\ref{hamiltonian_z}) RSOI removes the spin degeneracy of each energy branch and splits them into
two ones. The splitting is also accompanied by a deviation from the quadratic dependence on the momentum. For example in
the case of a harmonic $z$-confinement, if only the first two transverse sub-bands are considered, there are four 1D
dispersion relations $\varepsilon^{\text{1D}}_{\chi}(k_x)$, $\chi=1,2,3,4$. The eigen-states have a four-dimensional spinorial part
$\phi^{\text{1D}}_{k_x,\chi}(\sigma,j)$ where $j=0,1$ is the transverse mode index. It turns out that for this model the energy spectrum
can be found in analytic form \cite{Governale} from the diagonalization of Hamiltonian (\ref{hamiltonian_z}).

Structures where a periodic modulation $U(x)$ is additionally present have recently been investigated by various authors.
For the case $U(x)\neq0$, $V(z)=0$ the Bloch band energies have been found in Ref. \cite{Kleinert} within the tight-binding
approximation. In Ref. \cite{Demikhovskii} the same problem has been investigated numerically. In the presence of an
external homogeneous magnetic field the so-called magnetic Bloch states are discussed in Ref. \cite{Demikhovskii_1} for
the case $U(x)\neq0$ and $V(z)$ being periodic as well. However, analytic relations between the eigen-values of those
problems and the ones of their corresponding truly 1D problems without RSOI have not been provided so far and this is one
of the topics of the present letter.

In this work we consider two potentials $U(x)$ and $V(z)$, where the potential $U(x)$ is a periodic potential while the
shape of the potential $V(z)$ is arbitrary. First, a general structure of the Bloch amplitude is educed. Next, we
formulate the eigen-value problem. As an example we apply the general approach to the particular case of a harmonic
confinement and as a consequence generalize the analytical results obtained in Ref. \cite{Governale} to the case of a
periodic potential along the wire. Finally, setting $V(z)=0$ we analytically and {\it exactly solve} the problem examined
numerically in \cite{Demikhovskii} and find qualitative differences from some of the numerical results obtained
in \cite{Demikhovskii}.

\section{A periodic structure with RSOI}
In this section we consider a system described by the Hamiltonian $\hat{H}^{\text{2D}}_z+U(\hat{x})$ where the 1D periodic
potential $U(x)$ has the period $L$:
\begin{equation}
U(x+L)=U(x).
\end{equation}
That is, the total Hamiltonian of the problem is
\begin{equation}
\hat{H}=\frac{\hbar^2\hat{\vec{k}}^2}{2m}+V(\hat{z})-
\frac{\hbar^2k_{\text{so}}}{m}\bigl(\hat{\sigma}_{x}\hat{k}_z-\hat{\sigma}_z\hat{k}_x\bigl)+U(\hat{x}).
\label{hamiltonian}
\end{equation}
Before considering the full problem it is instructive to refresh the Bloch theorem for a truly 1D periodic structure
without RSOI.
\subsection{A truly 1D periodic structure}
As it is known \cite{AM}, a system described by the Hamiltonian
\begin{equation}
\hat{H}^{\text{1D}}_0=\frac{\hbar^2\hat{k}^2_x}{2m}+U(\hat{x})
\label{hamiltonian_1d_0}
\end{equation}
has eigen-energies $\varepsilon^{(0)}_{l,\sigma}(k_\text{B})$ and eigen-states $|l,k_\text{B},\sigma\rangle$, with
\begin{equation}
\hat{H}^{\text{1D}}_0|l,k_\text{B},\sigma\rangle=\varepsilon^{(0)}_{l,\sigma}(k_\text{B})|l,k_\text{B},\sigma\rangle,
\label{eigen_value_problem_1d_0}
\end{equation}
characterized by Bloch's quasi-momentum $k_\text{B}$, running over a discrete set of values in the first Brillouin zone, and
the band index $l$. The eigen-energies are degenerate with respect to the spin index,
$\varepsilon^{(0)}_{l,+1}(k_\text{B})=\varepsilon^{(0)}_{l,-1}(k_\text{B})\equiv\varepsilon^{(0)}_l(k_\text{B})$. In the coordinate representation the eigen-state is related to Bloch's
amplitude
$u_{l,k_\text{B},\sigma}(x,\sigma')$ by
\begin{equation}
\begin{split}
&\langle x,\sigma'|l,k_\text{B},\sigma\rangle=\frac{1}{\sqrt{L_0}}e^{\text{i}k_\text{B}x}u_{l,k_\text{B},\sigma}(x,\sigma'),\\
&u_{l,k_\text{B},\sigma}(x,\sigma')=\delta_{\sigma',\sigma}u_{l,k_\text{B}}(x),\\
&u_{l,k_\text{B}}(x)=u_{l,k_\text{B}}(x+L),
\end{split}
\label{Bloch_amplitude_1d_0}
\end{equation}
where $L_0$ is the size of the system along the $x$-axis. Here the spinorial structure of the Bloch amplitude is trivial.
\subsection{Influence of a transverse potential and RSOI}
The transverse potential $V(z)$ together with RSOI change the Bloch spinors $|l,k_\text{B},\sigma\rangle$. We denote the new spinors
through
$|l,k_\text{B},\eta\rangle$:
\begin{equation}
\hat{H}|l,k_\text{B},\eta\rangle=\varepsilon_{l,\eta}(k_\text{B})|l,k_\text{B},\eta\rangle.
\label{eigen_value_problem}
\end{equation}
As a result the Bloch amplitude acquires a new spinorial structure:
\begin{equation}
\begin{split}
&\langle x,j,\sigma|l,k_\text{B},\eta\rangle=\frac{1}{\sqrt{L_0}}e^{\text{i}k_\text{B}x}u_{l,k_\text{B},\eta}(x;j,\sigma),\\
&u_{l,k_\text{B},\eta}(x;j,\sigma)=u_{l,k_\text{B},\eta}(x+L;j,\sigma),
\end{split}
\label{spinoria_Bloch_amplitude_x_j_sigma}
\end{equation}
where $|j\rangle$ is an eigen-vector corresponding to an eigen-value $\varepsilon_j^z$ and both are found from the Schr\"odinger equation:
\begin{equation}
\biggl[\frac{\hbar^2\hat{k}^2_z}{2m}+V(\hat{z})\biggl]|j\rangle=\varepsilon_j^z|j\rangle.
\label{Schroedinger_z}
\end{equation}
It is convenient to represent the total Hamiltonian (\ref{hamiltonian}) as the sum $\hat{H}=\hat{H}'+\hat{H}''$, where
$\hat{H}'$ and $\hat{H}''$ are given by
\begin{equation}
\begin{split}
&\hat{H}'\equiv\frac{\hbar^2}{2m}\bigl(\hat{k}_x+\hat{\sigma}_zk_{\text{so}}\bigl)^2+U(\hat{x})+\\
&+\frac{\hbar^2\hat{k}_z^2}{2m}+V(\hat{z})-\frac{\hbar^2k^{2}_{\text{so}}}{2m},\\
&\hat{H}''\equiv-\frac{\hbar^2k_{\text{so}}}{m}\hat{\sigma}_x\hat{k}_z.
\end{split}
\label{H'_and_H''}
\end{equation}
The eigen-energies and eigen-states of $\hat{H}'$ are easily found and related to $\varepsilon^{(0)}_l(k_\text{B})$ and $u_{l,k_\text{B}}(x)$
as follows:
\begin{equation}
\begin{split}
&\hat{H}'|l,k_\text{B},j,\sigma\rangle=\varepsilon'_{l,j,\sigma}(k_\text{B})|l,k_\text{B},j,\sigma\rangle,\\
&\varepsilon'_{l,j,\sigma}(k_\text{B})=\varepsilon^{(0)}_l(k_\text{B}+\sigma k_{\text{so}})-\frac{\hbar^2k_{\text{so}}^2}{2m}+\varepsilon_j^z,\\
&\langle x,j',\sigma'|l,k_\text{B},j,\sigma\rangle=\frac{\delta_{j',j}\delta_{\sigma',\sigma}e^{\text{i}k_\text{B}x}}{\sqrt{L}_0}u_{l,k_\text{B}+\sigma k_{\text{so}}}(x).
\end{split}
\label{eigen_value_problem_H'}
\end{equation}
Let us denote through $\theta_{l,k_\text{B},\eta}(j,\sigma)$ the Bloch spinors in the $\{l,k_\text{B},j,\sigma \}$ representation, that is
$\theta_{l,k_\text{B},\eta}(j,\sigma)\equiv\langle l,k_\text{B},j,\sigma|l,k_\text{B},\eta\rangle$. Then
\begin{equation}
\langle l',k'_\text{B},j,\sigma|l,k_\text{B},\eta\rangle=\delta_{l',l}\delta_{k'_\text{B},k_\text{B}}\theta_{l,k_\text{B},\eta}(j,\sigma).
\label{spinorial_Bloch_amplitude_l_kB_j_sigma}
\end{equation}
We can make a general statement concerning Bloch's amplitude $u_{l,k_\text{B},\eta}(x;j,\sigma)$. From the identity
\begin{equation}
\begin{split}
&\frac{1}{\sqrt{L_0}}e^{\text{i}k_\text{B}x}u_{l,k_\text{B},\eta}(x;j,\sigma)\equiv\\
&\equiv\sum_{l',k'_\text{B},j',\sigma'}\langle x,j,\sigma|l',k'_\text{B},j',\sigma'\rangle\langle l',k'_\text{B},j',\sigma'|l,k_\text{B},\eta\rangle
\end{split}
\label{auxiliary_identity}
\end{equation}
and using (\ref{eigen_value_problem_H'}) and (\ref{spinorial_Bloch_amplitude_l_kB_j_sigma}) it follows
\begin{equation}
u_{l,k_\text{B},\eta}(x;j,\sigma)=u_{l,k_\text{B}+\sigma k_{\text{so}}}(x)\theta_{l,k_\text{B},\eta}(j,\sigma).
\label{relation_between_Bloch_amplitudes}
\end{equation}
The last equation clearly shows that the spinorial part of the Bloch amplitude $u_{l,k_\text{B},\sigma}(x,\sigma')$ in eq.
(\ref{Bloch_amplitude_1d_0}) transforms from $\delta_{\sigma',\sigma}$ into the spinor $\theta_{l,k_\text{B},\eta}(j,\sigma)$ when the potential $V(z)$ and
RSOI are involved.

The spinors $\theta_{l,k_\text{B},\eta}(j,\sigma)$ can be found from eq. (\ref{eigen_value_problem}) which in $\{l,k_\text{B},j,\sigma \}$
representation takes the form:
\begin{equation}
\begin{split}
&\sum_{j',\sigma'}\bigl[\langle l',k'_\text{B},j,\sigma|\hat{H}'|l,k_\text{B},j',\sigma'\rangle+\\
&+\langle l',k'_\text{B},j,\sigma|\hat{H}''|l,k_\text{B},j',\sigma'\rangle\bigl]\theta_{l,k_\text{B},\eta}(j',\sigma')=\\
&=\varepsilon_{l,\eta}(k_\text{B})\delta_{l',l}\delta_{k'_\text{B},k_\text{B}}\theta_{l,k_\text{B},\eta}(j,\sigma).
\end{split}
\label{eigen_value_problem_1}
\end{equation}
The matrix elements of $\hat{H}'$ and $\hat{H}''$ are given by the expressions:
\begin{equation}
\begin{split}
&\langle l',k'_\text{B},j,\sigma|\hat{H}'|l,k_\text{B},j',\sigma'\rangle=\\
&=\varepsilon'_{l,j,\sigma}(k_\text{B})\delta_{l',l}\delta_{k'_\text{B},k_\text{B}}\delta_{j,j'}\delta_{\sigma,\sigma'},\\
&\langle l',k'_\text{B},j,\sigma|\hat{H}''|l,k_\text{B},j',\sigma'\rangle=\\
&=-\frac{\hbar^2k_{\text{so}}}{m}\langle\sigma|\hat{\sigma}_x|\sigma'\rangle\langle j|\hat{k}_z|j'\rangle\delta_{l',l}\delta_{k'_\text{B},k_\text{B}},
\end{split}
\label{matrix_elements_H'_and_H''}
\end{equation}
where $\langle\sigma|\hat{\sigma}_x|\sigma'\rangle=1-\delta_{\sigma,\sigma'}$. The final equation for the eigen-energies $\varepsilon_{l,\eta}(k_\text{B})$ and eigen-spinors
$\theta_{l,k_\text{B},\eta}(j,\sigma)$ is obtained using eq. (\ref{eigen_value_problem_1}) together with eq.
(\ref{matrix_elements_H'_and_H''}) by equating the band indices $l'=l$ and Bloch's quasi-momenta $k'_\text{B}=k_\text{B}$:
\begin{equation}
\begin{split}
&\sum_{j',\sigma'}\biggl\{\delta_{j,j'}\delta_{\sigma,\sigma'}\biggl[\varepsilon^{(0)}_l(k_\text{B}+\sigma k_{\text{so}})+\varepsilon_j^z-\frac{\hbar^2k_{\text{so}}^2}{2m}\biggl]-\\
&-\frac{\hbar^2k_{\text{so}}}{m}(1-\delta_{\sigma,\sigma'})\langle j|\hat{k}_z|j'\rangle\biggl\}\theta_{l,k_\text{B},\eta}(j',\sigma')=\\
&=\varepsilon_{l,\eta}(k_\text{B})\theta_{l,k_\text{B},\eta}(j,\sigma).
\end{split}
\label{eigen_value_problem_final}
\end{equation}
We want to emphasize that eq. (\ref{eigen_value_problem_final}) can be applied to calculate the band structure for an
arbitrary potential $V(z)$ and periodic potential $U(x)$ of a general form. Note the specific influence of the spin-orbit
coupling: a) $j$-states are mixed by RSOI; b) the Bloch bands of the corresponding truly 1D problem with
different $l$ are split into sub-bands independently, that is the splitting of band $l$ {\it does not depend} on the
splitting of bands with $l'\neq l$. Therefore as soon as the truly 1D band structure has been obtained, one can take any of
its Bloch bands, let us say $l$, apply (\ref{eigen_value_problem_final}) to it and find the Bloch bands labeled with index
$l$ in the presence of $V(z)$ and RSOI. The same inference remains valid if Dresselhaus' spin-orbit interaction is
additionally included into the model.

\section{Harmonic confinement}
\begin{figure}
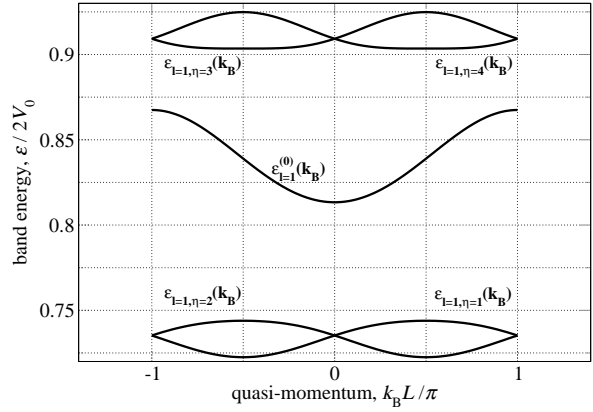

\onefigure{fig1.eps}
\caption{The first Bloch band of the corresponding truly 1D system $\varepsilon^{(0)}_{l=1}(k_\text{B})$ together with the four Bloch
  sub-bands $\varepsilon_{l=1,\eta}(k_\text{B})$ of the quasi-1D system in the presence of RSOI and the transverse confinement.}
\label{fig.1}
\end{figure}

\begin{figure}
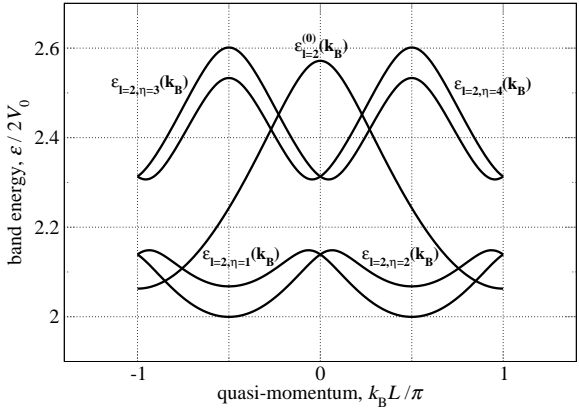

\onefigure{fig2.eps}
\caption{The second Bloch band of the corresponding truly 1D system $\varepsilon^{(0)}_{l=2}(k_\text{B})$ together with the four Bloch
  sub-bands $\varepsilon_{l=2,\eta}(k_\text{B})$ of the quasi-1D system in the presence of RSOI and the transverse confinement.}
\label{fig.2}
\end{figure}

Here we consider a particular case where the operator $V(\hat{z})$ represents a harmonic confinement of strength $\omega_0$. In
this case the matrix elements of $\hat{k}_z$ are
\begin{equation}
\langle j|\hat{k}_z|j'\rangle=\pm \text{i}\,\delta_{j,j'\pm1}\sqrt{\frac{(j+\frac{1}{2}\mp\frac{1}{2})m\omega_0}{2\hbar}}.
\end{equation}
If in (\ref{eigen_value_problem_final}) one keeps only the first two transverse modes, the problem reduces to the
diagonalization of a $4\times4$ matrix and becomes solvable analytically. The validity of this approximation is discussed in
Ref. \cite{Governale}. After the diagonalization of (\ref{eigen_value_problem_final}), where now $j=0,1$, we obtain the
following eigen-energies:
\begin{equation}
\begin{split}
&\varepsilon_{l,\eta=1,2}(k_\text{B})=\varepsilon_l^+(k_\text{B})-\Xi_{l_{1,2}}(k_\text{B}),\\
&\varepsilon_{l,\eta=3,4}(k_\text{B})=\varepsilon_l^+(k_\text{B})+\Xi_{l_{2,1}}(k_\text{B}),
\end{split}
\label{eigen_values}
\end{equation}
where
\begin{equation}
\begin{split}
&\varepsilon_l^+(k_\text{B})\equiv\frac{\varepsilon^{(0)}_l(k_\text{B}+k_{\text{so}})+\varepsilon^{(0)}_l(k_\text{B}-k_{\text{so}})}{2}+\\
&+\hbar\omega_0-\frac{\hbar^2k_{\text{so}}^2}{2m},\\
&\Xi_{l_{1,2}}(k_\text{B})\equiv\sqrt{\Xi^2+\biggl(\varepsilon_l^-(k_\text{B})\mp\frac{\hbar\omega_0}{2}\biggl)^2},\\
&\varepsilon_l^-(k_\text{B})\equiv\frac{\varepsilon^{(0)}_l(k_\text{B}+k_{\text{so}})-\varepsilon^{(0)}_l(k_\text{B}-k_{\text{so}})}{2},\\
&\Xi\equiv\frac{\hbar^2k_{\text{so}}}{m}\sqrt{\frac{m\omega_0}{2\hbar}}.
\end{split}
\label{epsilon_plus_minus}
\end{equation}
Since $\varepsilon^{(0)}_l(k_\text{B})=\varepsilon^{(0)}_l(-k_\text{B})$, the relations between the eigen-energies (\ref{eigen_values}) follow:
\begin{equation}
\begin{split}
&\varepsilon_{l,\eta=1}(k_\text{B})=\varepsilon_{l,\eta=2}(-k_\text{B}),\\
&\varepsilon_{l,\eta=3}(k_\text{B})=\varepsilon_{l,\eta=4}(-k_\text{B}),
\end{split}
\label{relations_between_eigen_energies}
\end{equation}
as expected due to the existence of both the time reversal symmetry and band overlap \cite{LL_IX}. In fig.~\ref{fig.1} we
show the first Bloch band of the corresponding truly 1D problem and the four Bloch sub-bands growing out of it under the
influence of RSOI and the transverse confinement. The spin-orbit coupling strength is chosen such that $Lk_{\text{so}}=\pi/2$. The
periodic potential has the form:
\begin{equation}
U(x)=V_0\biggl[1-\cos\biggl(\frac{2\pi}{L}x\biggl)\biggl].
\end{equation}
The second Bloch band and its four sub-bands are plotted in fig.~\ref{fig.2}. It can be seen that for
$l=1$ the Bloch band of the truly 1D problem without RSOI and its four sub-bands for the quasi-1D system with RSOI are all
under the potential barrier while for $l=2$ they are above it. As usual RSOI does not remove the spin degeneracy at
$k_\text{B}=0$. It follows from (\ref{eigen_values}) that the bands split when $\varepsilon^-_l(k_\text{B})\neq0$. The derivative of the function
$\varepsilon^-_l(k_\text{B})$ at $k_\text{B}=0$ is easily found from (\ref{epsilon_plus_minus}):
\begin{equation}
\frac{d\varepsilon^-_l(k_\text{B})}{dk_\text{B}}\biggl|_{k_\text{B}=0}= v_l^{(0)}(k_{\text{so}}),
\label{epsilon_minus_approx}
\end{equation}
where $v_l^{(0)}(k_\text{B})$ is the group velocity of the corresponding truly 1D problem. Since for the chosen parameters the
group velocity in (\ref{epsilon_minus_approx}) is not equal to zero (see figs.~\ref{fig.1} and \ref{fig.2}), it follows
from (\ref{eigen_values}) and (\ref{epsilon_plus_minus}) that the band splitting near the point $k_\text{B}=0$ is
{\it linear} in $k_\text{B}$. This is also the case for a 2DEG where the linear momentum-dependence of the splitting is
observed experimentally \cite{Luo}.

The corresponding normalized eigen-spinors $\theta_{l,k_\text{B},\eta}(j,\sigma)$ are expressed in terms of non-normalized ones, denoted
through
$\tilde{\theta}_{l,k_\text{B},\eta}(j,\sigma)$, as:
\begin{equation}
\begin{split}
&\theta_{l,k_\text{B},\eta=1,4}=N^{-\frac{1}{2}}_{l,k_\text{B},\eta=1,4}\tilde{\theta}_{l,k_\text{B},\eta=1,4}\,,\\
&\theta_{l,k_\text{B},\eta=2,3}=N^{-\frac{1}{2}}_{l,k_\text{B},\eta=2,3}\tilde{\theta}_{l,k_\text{B},\eta=2,3}\,,\\
\end{split}
\label{eigen_spinors}
\end{equation}
where
\begin{equation}
\begin{split}
&\tilde{\theta}_{l,k_\text{B},\eta=1,4}\equiv\begin{bmatrix}
  \frac{\text{i}}{\Xi}\biggl[\varepsilon_l^-(k_\text{B})-\frac{\hbar\omega_0}{2}\mp\Xi_{l_1}(k_\text{B})\biggl]\\
  0\\
  0\\
  1\\
\end{bmatrix},\\
&\tilde{\theta}_{l,k_\text{B},\eta=2,3}\equiv\begin{bmatrix}
  0\\
  -\frac{\text{i}}{\Xi}\biggl[\varepsilon_l^-(k_\text{B})+\frac{\hbar\omega_0}{2}\pm\Xi_{l_2}(k_\text{B})\biggl]\\
  1\\
  0\\
\end{bmatrix}.
\end{split}
\label{eigen_spinors_non_norm}
\end{equation}
We have introduced the notation
\begin{equation}
  \theta_{l,k_\text{B},\eta}\equiv\begin{bmatrix}
  \theta_{l,k_\text{B},\eta}(j=0,\sigma=+1)\\
  \theta_{l,k_\text{B},\eta}(j=0,\sigma=-1)\\
  \theta_{l,k_\text{B},\eta}(j=1,\sigma=+1)\\
  \theta_{l,k_\text{B},\eta}(j=1,\sigma=-1)\\
\end{bmatrix},
\label{spinorial_notations}
\end{equation}
and an analogous one for the non-normalized spinor $\tilde{\theta}_{l,k_\text{B},\eta}$. In (\ref{eigen_spinors}) $N_{l,k_\text{B},\eta}$
are the normalization constants:
\begin{equation}
N_{l,k_\text{B},\eta}=\sum_{j=0}^1\sum_{\sigma=-1}^{+1}|\tilde{\theta}_{l,k_\text{B},\eta}(j,\sigma)|^2.
\label{normalization_constant}
\end{equation}
Note that using (\ref{epsilon_plus_minus}) and (\ref{eigen_spinors}) one gets the relations
\begin{equation}
N_{l,k_\text{B},\eta=1,4}=N_{l,-k_\text{B},\eta=2,3}.
\label{normalization_constant_relations}
\end{equation}
For $k_{\text{so}}\to0$ the spinors in (\ref{eigen_spinors}) take the form:
\begin{equation}
\theta_{l,k_\text{B},\eta=1,2,3,4}=\begin{bmatrix}
  -\text{i}\\
  0\\
  0\\
  0\\
\end{bmatrix},
\begin{bmatrix}
  0\\
  -\text{i}\\
  0\\
  0\\
\end{bmatrix},
\begin{bmatrix}
  0\\
  0\\
  1\\
  0\\
\end{bmatrix},
\begin{bmatrix}
  0\\
  0\\
  0\\
  1\\
\end{bmatrix}.
\label{eigen_spinors_limit_kso}
\end{equation}

In the limit $U(x)\to0$ we have $\varepsilon^{(0)}_l(k_\text{B})\to\hbar^2k_{B}^2/2m$, $u_{l,k_\text{B}}(x)\to1$ and from (\ref{eigen_values}) and
(\ref{epsilon_plus_minus}) we get:
\begin{equation}
\begin{split}
&\varepsilon_{l,\eta=1,2}(k_x)\to\frac{\hbar^2k^2_x}{2m}+\hbar\omega_0-\Xi_{1,2}^{(0)}(k_x),\\
&\varepsilon_{l,\eta=3,4}(k_x)\to\frac{\hbar^2k^2_x}{2m}+\hbar\omega_0+\Xi_{2,1}^{(0)}(k_x),
\end{split}
\label{energy_values_limit}
\end{equation}
where
\begin{equation}
\Xi_{1,2}^{(0)}(k_x)\equiv\sqrt{\Xi^2+\biggl(\frac{\hbar^2k_xk_{\text{so}}}{m}\mp\frac{\hbar\omega_0}{2}\biggl)^2}.
\label{Xi_12_0}
\end{equation}
Further, in this limit from (\ref{eigen_spinors_non_norm}) we find:
\begin{equation}
\begin{split}
&\tilde{\theta}_{l,k_x,\eta=1,4}\to \begin{bmatrix}
  \frac{\text{i}}{\Xi}\biggl[\frac{\hbar^2k_xk_{\text{so}}}{m}-\frac{\hbar\omega_0}{2}\mp\Xi_1^{(0)}(k_x)\biggl]\\
  0\\
  0\\
  1\\
\end{bmatrix},\\
&\tilde{\theta}_{l,k_x,\eta=2,3}\to\begin{bmatrix}
  0\\
  -\frac{\text{i}}{\Xi}\biggl[\frac{\hbar^2k_xk_{\text{so}}}{m}+\frac{\hbar\omega_0}{2}\pm\Xi_2^{(0)}(k_x)\biggl]\\
  1\\
  0\\
\end{bmatrix}.
\end{split}
\label{eigen_spinors_limit}
\end{equation}
As a consequence the spinorial Bloch amplitude transforms into a pure spinor without any real space dependence as it can
be seen from (\ref{relation_between_Bloch_amplitudes}). Expressions (\ref{energy_values_limit}) and (\ref{Xi_12_0})
recover the results obtained in Ref. \cite{Governale}.

Finally, let us discuss the polarizations
\begin{equation}
P^{(i)}_{l,\eta}(k_\text{B})\equiv\langle l,k_\text{B},\eta|\hat{\sigma}_i|l,k_\text{B},\eta\rangle,
\label{polarization}
\end{equation}
where $\hat{\sigma}_i$, $i=x,y,z$ are the Pauli spin operators. Writing the identity operator in the $\{|l,k_\text{B},j,\sigma\rangle\}$ basis,
and taking into account the structure of the Bloch spinors (\ref{eigen_spinors}) and (\ref{eigen_spinors_non_norm}) we
obtain
\begin{equation}
\begin{split}
&P_{l,\eta}^{(x)}(k_\text{B})=\sum_{j=0}^1\sum_{\sigma',\sigma''=-1}^{+1}\bigl[\theta^*_{l,k_\text{B},\eta}(j,\sigma')\times\\
&\times(1-\delta_{\sigma',\sigma''})\theta_{l,k_\text{B},\eta}(j,\sigma'')\bigl]=0,\quad\forall\,l,k_\text{B}\in\text{B.Z.},
\end{split}
\label{polarization_x}
\end{equation}
\begin{equation}
\begin{split}
&P_{l,\eta}^{(y)}(k_\text{B})=\sum_{j=0}^1\sum_{\sigma',\sigma''=-1}^{+1}\bigl[\theta^*_{l,k_\text{B},\eta}(j,\sigma')\,\text{i}^{\sigma''}\times\\
&\times(1-\delta_{\sigma',\sigma''})\theta_{l,k_\text{B},\eta}(j,\sigma'')\bigl]=0,\quad\forall\,l,k_\text{B}\in\text{B.Z.},
\end{split}
\label{polarization_y}
\end{equation}
where $\eta=1,2,3,4$. The last two equations show that the longitudinal, that is along the wire, and the perpendicular to the
2DEG plane components of the polarization identically vanish. However, the polarization along the in-plane confinement
direction has a finite value:
\begin{equation}
\begin{split}
&P_{l,\eta}^{(z)}(k_\text{B})=\sum_{j=0}^{1}\sum_{\sigma=-1}^{+1}\theta^*_{l,k_\text{B},\eta}(j,\sigma)\,\sigma\,\theta_{l,k_\text{B},\eta}(j,\sigma),\\
&\quad\forall\,l,k_\text{B}\in\text{B.Z.},\eta=1,2,3,4.
\end{split}
\label{polarization_z}
\end{equation}
From eqs. (\ref{polarization_z}) and (\ref{eigen_spinors}) we derive the polarizations in the four Bloch sub-bands formed
out of the truly 1D Bloch band with index $l$:
\begin{equation}
\begin{split}
&P_{l,\eta=1,4}^{(z)}(k_\text{B})=N_{l,k_\text{B},\eta=1,4}^{-1}\times\\
&\times\biggl\{\frac{1}{\Xi^2}\biggl[\varepsilon_l^-(k_\text{B})-\frac{\hbar\omega_0}{2}\mp\Xi_{l_1}(k_\text{B})\biggl]^2-1\biggl\},
\end{split}
\label{polarization_z_1}
\end{equation}
\begin{equation}
\begin{split}
&P_{l,\eta=2,3}^{(z)}(k_\text{B})=N_{l,k_\text{B},\eta=2,3}^{-1}\times\\
&\times\biggl\{1-\frac{1}{\Xi^2}\biggl[\varepsilon_l^-(k_\text{B})+\frac{\hbar\omega_0}{2}\pm\Xi_{l_2}(k_\text{B})\biggl]^2\biggl\}.
\end{split}
\label{polarization_z_2}
\end{equation}
Using (\ref{normalization_constant_relations}) and equalities $\varepsilon_l^-(k_\text{B})=-\varepsilon_l^-(-k_\text{B})$,
$\Xi_{l_{1,2}}(k_\text{B})=\Xi_{l_{2,1}}(-k_\text{B})$, the symmetry relation for the polarizations
\begin{equation}
P_{l,\eta=1,4}^{(z)}(k_\text{B})=-P_{l,\eta=2,3}^{(z)}(-k_\text{B}),
\label{relations_between_polarizations_z}
\end{equation}
is derived $\forall\,l,k_\text{B}\in \text{B.Z.}$ This symmetry is clearly seen in figs.~\ref{fig.3} and~\ref{fig.4}, where the four
polarizations
\begin{figure}
\onefigure{fig3.eps}
\caption{Spin polarizations along the $z$-axis in the four Bloch sub-bands with $l=1$.}
\label{fig.3}
\end{figure}
\begin{figure}
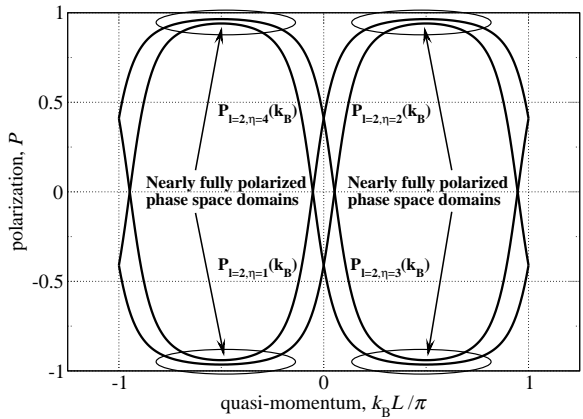

\onefigure{fig4.eps}
\caption{Spin polarizations along the $z$-axis in the four Bloch sub-bands with $l=2$.}
\label{fig.4}
\end{figure}
(\ref{polarization_z_1}) and (\ref{polarization_z_2}) are plotted for $l=1$ and $l=2$, respectively. Fig.~\ref{fig.3} also
shows that the polarizations do not change sign and never approach unity in the first Bloch band. This picture changes for
the polarizations in the second Bloch band (see fig.~\ref{fig.4}). In this band the polarizations change sign. Also there
exist nearly fully spin-polarized domains in the first Brillouin zone. As one can see those domains are the ones where the
group velocity takes its largest absolute values. The same happens in the limiting case $U(x)\to0$ where the group velocity
has infinite values for infinite momentum. Indeed, when $U(x)\to0$, from (\ref{polarization_z_1}) one finds for example that
$\underset{k_x\to\pm\infty}{\lim}P_{\eta=1}^{(z)}(k_x)=\mp1$ in agreement with Ref.\cite{Governale}. Thus in the absence of the periodic
potential the states can again be characterized by the spin quantum number for large absolute values of the longitudinal
momentum.

\section{A periodic structure with $V(z)=0$}
\begin{figure}
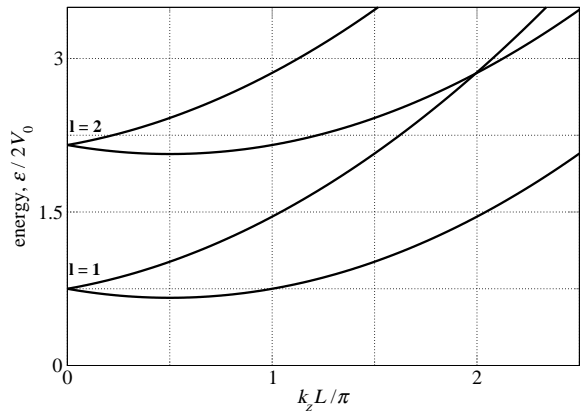

\onefigure{fig5.eps}
\caption{$k_\text{B}=0$ cut of the energy spectrum for a periodic structure with $V(z)=0$. Energy branches with $l=1,2$
  are depicted.}
\label{fig.5}
\end{figure}

In this section we briefly present the resulting energy spectrum when the potential $V(z)$ vanishes and the periodic
potential $U(x)$ is arbitrary. Here the solutions of (\ref{Schroedinger_z}) are plane waves, $|j\rangle\equiv|k_z\rangle$,
$\varepsilon_j^z\equiv\varepsilon_{k_z}^z=\hbar^2k_z^2/2m$ and $\langle k_z|\hat{k}_z|k'_z\rangle=\delta_{k_z,k'_z}k_z$. The diagonalization of eq. (\ref{eigen_value_problem_final})
leads to the dispersion relations:
\begin{equation}
\begin{split}
&\varepsilon^{\text{2D}}_{l,\eta=1,2}(k_\text{B},k_z)=\frac{\varepsilon_l^{(0)}(k_\text{B}+k_{\text{so}})+\varepsilon_l^{(0)}(k_\text{B}-k_{\text{so}})}{2}+\\
&+\frac{\hbar^2k_z^2}{2m}-\frac{\hbar^2k_{\text{so}}^2}{2m}\pm\sqrt{\biggl(\varepsilon_l^-(k_\text{B})\biggl)^2+\biggl(\frac{\hbar^2k_{\text{so}}k_z}{m}\biggl)^2},
\end{split}
\label{eigen_value_Vz0}
\end{equation}
where we have added the upper index 2D to stress that in this system the energy spectrum is two-dimensional. It can be
easily checked that at $k_z=0$ eq. (\ref{eigen_value_Vz0}) gives the same dispersion relation as the one derived from eqs.
(\ref{eigen_values}) and (\ref{epsilon_plus_minus}) in the limiting case $\omega_0=0$. For $k_\text{B}=0$ and $k_z>0$ it
follows from (\ref{eigen_value_Vz0}):
\begin{equation}
\begin{split}
&\varepsilon^{\text{2D}}_{l,\eta=1,2}(k_\text{B}=0,k_z)=\varepsilon_l^{(0)}(k_{\text{so}})-\frac{\hbar^2k_{\text{so}}^2}{2m}+\\
&+\frac{\hbar^2k_z^2}{2m}\pm\frac{\hbar^2k_{\text{so}}k_z}{m}.
\end{split}
\label{eigen_value_Vz0_kb0}
\end{equation}
From eq. (\ref{eigen_value_Vz0_kb0}) one can clearly see that the energy branch with $\eta=2$ has its minimum at $k_z=k_{\text{so}}$
for all bands $l$. The splitting of the two branches is linear in $k_z$. The last expression also shows that for different
band indices $l$ the corresponding energy branches are parallel and there are not anti-crossings. This is also shown in
fig.~\ref{fig.5}
These analytical results are in contrast to what was numerically predicted in Ref. \cite{Demikhovskii}.

Finally, in the limiting case $U(x)\to0$, we have $\varepsilon_l^{(0)}(k_\text{B})\to\hbar^2k_x^2/2m$ and from eq. (\ref{eigen_value_Vz0}) it follows:
\begin{equation}
\varepsilon_{1,2}^{\text{2D}}(k_x,k_z)=\frac{\hbar^2k^2}{2m}\pm\frac{\hbar^2k_{\text{so}}k}{m},
\label{Rashba_dispersion}
\end{equation}
where $k\equiv|\vec{k}|=\sqrt{k_x^2+k_z^2}$. One sees that eq. (\ref{Rashba_dispersion}) is nothing but Rashba's dispersion relation,
that is the energy spectrum of Hamiltonian (\ref{hamiltonian_2D}) has been recovered.

\section{Materials of interest}
Here we would like to mention that although our theory is general, the concrete results presented on the plots are
relevant for III-V compounds. For example in InAs the spin-orbit coupling strength $\alpha\equiv\hbar^2k_\text{so}/m$ is enhanced up to
$4\cdot10^{-11}\,\text{eV}\cdot\text{m}$ as it is demonstrated in Ref. \cite{Grundler}. The effective mass is $m=0.036m_0$. Then for
$L$ in the range between 70 nm and 100 nm the dimensionless parameter $k_\text{so}L=g\pi/2$ with $g$ being in the range between
0.84 and 1.2.

\section{Conclusion}
We have considered a two-dimensional (2D) electron gas with Rashba's spin-orbit interaction (RSOI) in the presence of two
one-dimensional (1D) in-plane potentials along mutually orthogonal directions, assuming the first of those potentials to
be periodic while making no assumption about the second one. It has been found that in such a system the coordinate part
of the Bloch amplitude is the same as the one of the corresponding truly one-dimensional problem without RSOI, however its
Bloch's quasi-momentum has a spin-dependent shift proportional to the spin-orbit coupling strength. A general eigen-value
problem for the band structure has been presented in terms of the spinorial part of Bloch's amplitude. The cases where the
second potential represents either a harmonic confinement or where it vanishes have been studied as applications of the
general formalism. For the case of a harmonic confinement with only the first two transverse modes retained
{\it analytical} relations have been obtained and general symmetry properties of the resulting band structure have been
determined. Analytical expressions for the polarizations have been derived as well. Regions of high polarization
corresponding to regions of large absolute values of the group velocity have been found. For a vanishing transverse
potential {\it exact analytical relations} between the energy spectrum of this 2D system and its truly 1D problem
without RSOI have also been established. We hope that the results of our work could be important to better understand the
interplay between RSOI and periodic potentials in a wide range of 2D and quasi-1D systems which could be used {\it e.g.}
as effective spin rectifiers \cite{Scheid}.

\acknowledgments
We thank Prof. K. Richter for fruitful discussions. Support from SFB 689 is acknowledged.


\begin{thebibliography}{0}

\bibitem{Rashba}
  \Name{Rashba E.}
  \REVIEW{Fiz. Tverd. Tela (Leningrad)}{2}{1960}{1224},
  \REVIEW{Sov. Phys. Solid State}{2}{1960}{1109}.

\bibitem{Nitta}
  \Name{Nitta J., Akazaki T. \and Takayanagi H.}
  \REVIEW{Phys. Rev. Lett.}{78}{1997}{1335}.

\bibitem{Dresselhaus}
  \Name{Dresselhaus G.}
  \REVIEW{Phys. Rev.}{100}{1955}{580}.

\bibitem{Datta}
  \Name{Datta S. \and Das B.}
  \REVIEW{Appl. Phys. Lett.}{56}{1990}{665}.

\bibitem{Moroz}
  \Name{Moroz A.V. \and Barnes C.H.W.}
  \REVIEW{Phys. Rev. B}{60}{1999}{14272}; ibid. {\bf 61} (2000) R2464.

\bibitem{Governale}
  \Name{Governale M. \and Z\"ulicke U.}
  \REVIEW{Phys. Rev. B}{66}{2002}{073311}.

\bibitem{Perroni}
  \Name{Perroni C.A., Bercioux D., Ramaglia V.M. \and Cataudella V.}
  \REVIEW{J.Phys.: Condens. Matter}{19}{2007}{186227}.

\bibitem{Kleinert}
  \Name{Kleinert P., Bryksin V.V. \and Bleibaum O.}
  \REVIEW{Phys. Rev. B}{72}{2005}{195311}.

\bibitem{Demikhovskii}
  \Name{Demikhovskii V.Ya. \and Khomitsky D.V.}
  \REVIEW{Pis'ma Zh. Eksp. Teor. Fiz.}{83}{2006}{399}.

\bibitem{Demikhovskii_1}
  \Name{Demikhovskii V.Ya. \and Perov A.A.}
  \REVIEW{Europhys. Lett.}{76}{2006}{477}.

\bibitem{AM}
  \Name{Ashkroft N.W. \and Mermin N.D.}
  \Book{Solid State Physics}
  \Publ{Saunders College, Philadelphia}
  \Year{1976}.

\bibitem{LL_IX}
  \Name{Landau L.D., Lifshitz E.M. \and Pitaevskii L.P.}
  \Book{Course of Theoretical Physics, Statistical Physics. Part 2: Theory of the condensed state}
  \Vol{9}
  \Publ{Butterworth-Heinemann, Oxford}
  \Year{2002}.

\bibitem{Luo}
  \Name{Luo J., Munekata H., Fang F.F. \and Stiles P.J.}
  \REVIEW{Phys. Rev. B}{41}{1990}{7685}.

\bibitem{Grundler}
  \Name{Grundler D.}
  \REVIEW{Phys. Rev. Lett.}{84}{2000}{6074}.

\bibitem{Scheid}
  \Name{Scheid M., Pfund A., Bercioux D. \and Richter K.} {\it arXiv: cond-mat}/{\bf 0601118}.

\end{thebibliography}
\end{document}